\documentclass[reprint,showpacs,amsfonts,amsmath,amssymb,prl]{revtex4-1}
\usepackage{graphicx}
\usepackage{dcolumn}
\usepackage{bm}

\begin{document}

\title{Consistent picture of the octet-nodal gap and its evolution with doping in heavily overdoped Ba$_{1-x}$K$_{x}$Fe$_{2}$As$_{2}$}

\author{Shun-Li Yu}
\author{Zi-Jian Yao}
\author{Jian-Xin Li}
\affiliation{National Laboratory of Solid State Microstructures
and Department of Physics, Nanjing University, Nanjing 210093,
China\\
Collaborative Innovation Center of Advanced Microstructures,
Nanjing, China}

\date{\today}

\begin{abstract}
We investigate the pairing symmetry in heavily overdoped
Ba$_{1-x}$K$_{x}$Fe$_{2}$As$_{2}$ based on the spin-fluctuation
mechanism. The exotic octet nodes of the superconducting gap and
the unusual evolution of the gap with doping observed by the
recent experiments are well explained in a unified manner. We
demonstrate that the scatterings of electrons on the Fermi
patches is mainly responsible for the incommensurate spin
fluctuations and consequently the Fermi-surface-dependent
multi-gap structure, since the Fermi level is close to the flat
band. In addition, we find that a $d$-wave pairing state will
prevail over the s-wave pairing state around the Lifshitz
transition point.
\end{abstract}


\maketitle

Since the discovery of iron-based superconductors (FeSCs) in 2008
\cite{Kamihara,ChenXH}, the mechanism underlying the
superconductivity has been one of the most challenging problems.
Despite great efforts, the pairing mechanism and the pairing
symmetries are still under debate. The prevailing theoretical
suggestion is that the electron pairing is mediated by the
collinear antiferromagnetic spin-fluctuations and the
superconducting (SC) gap changes sign between the hole and
electron Fermi surfaces (FSs) (the so-called $s_{\pm}$ state) in
the underdoped and optimally doped
regimes\cite{IIMazin,KKuroki,ZJYao,FWang,SGraser},
though there are other proposals\cite{HKontani,KJSeo,QMSi}.
Recently, the over-hole-doped compounds Ba$_{1-x}$K$_{x}$Fe$_{2}$As$_{2}$
(BaK122) have attracted much attention, as they exhibit many
anomalies deviating from the known FeSC trends. Thus, the
understanding of the gap symmetries in these superconducting
materials will provide more insights into the microscopic pairing
mechanism in the FeSC.

KFe$_{2}$As$_{2}$ is the end member of the
BaK122 series with $x=1$. Unlike the
optimally doped systems ($x=0.4$), where the sizes of the electron
and hole FSs are roughly equal, KFe$_{2}$As$_{2}$ has only the
hole pockets centered at $\Gamma$ point [$\bm{k}=(0,0)$] but no
electron pockets according to the angle-resolved photoemission
spectroscopy (ARPES) measurements \cite{KOkazaki,TYoshida,TSato}.
The absence of the electron pockets makes the $s_{\pm}$
superconducting mechanism, which is mediated by the interband spin
fluctuations between the hole and electron pockets,
questionable in these heavily overdoped systems. A functional
renormalization group study \cite{RThomale} suggests a $d$-wave
superconducting state, while a random-phase-approximation (RPA)
analysis \cite{KSuzuki} shows that the $s_{\pm}$-wave pairing is
dominant but the $d$-wave is very close in energy.
Experimentally, the thermal conductivity
\cite{JKDong,JPhReid,AFWang} and magnetic penetration-depth
\cite{KHashimoto2} measurements support the $d$-wave symmetry.
However, the recent laser ARPES experiment observes a nodal
$s$-wave state with an exotic FS-dependent multi-gap structure
\cite{KOkazaki}: a nodeless gap with the largest magnitude on the
inner FS, an unconventional gap with ¡°octet-line nodes¡± on the
middle FS, and an almost-zero gap on the outer FS. These ARPES results
are supported by thermodynamic experiments
\cite{FHardy,DWatanabe}.
Moreover, the more recent ARPES experiment \cite{YOta} finds
that the gap anisotropy on these FSs drastically changes with a
small amount of electron doping. In particular, the gap on the
middle FS becomes nodeless when the electron concentration is
slightly increased, while the gap structure on the outer FS remains
unchanged\cite{YOta}. By assuming the dominant interaction at small
momentum transfers, a particular kind of $s$-wave state was
proposed \cite{SMaiti}, but it is hard to explain the evolution of gap structure with doping.
Besides the multi-gap structures, the spin fluctuation also
exhibits differences from that in optimally doped systems, where
spin fluctuations are observed at the same $q$ position $(\pi,0)$ as the collinear SDW order
wavevector at low energies\cite{Dai}. By contrast, the spin
fluctuation in overdoped BaK122 is
incommensurate with the wavevector $(\pi\pm0.32\pi,0)$ as found by
the recent inelastic neutron scattering (INS) experiments\cite{CHLee,JPCastellan}.

These anomalous phenomena in heavily overdoped BaK122 raise the
following issues: (i) How can we understand the complicated octet-node
structures of the SC gap and its evolution with doping? (ii) What
is the possible origin of the incommensurate spin fluctuations?
 (iii) Whether is the spin fluctuation mechanism applicable to
these systems in which the electron pockets are absent. In this
Letter, we provide a unified picture for the incommensurate spin
fluctuations and the resulting unusual octet-node structures of
the SC gap and its evolution with doping based on a weak-coupling
calculation of the five-orbital Hubbard model. We find that the
incommensurate spin fluctuation originates from the intra-orbital
particle-hole excitations. Though the inter-orbital spin
fluctuation does not show up in the physical spin susceptibility,
it plays a role in the determination of the FS-dependent exotic SC
gap structure. Besides the FS nesting, we find that the
scatterings of electrons related to the Fermi patches play an
essential role both in the incommensurate spin fluctuation and SC
gap structures, since the Fermi level is close to the flat band.
In addition, although the $s$-wave pairing is the dominant pairing
symmetry in a large doping range, a $d$-wave pairing state will
be more favored around the Lifshitz
transition point. Therefore, the long sought $s+id$ state is
probably realized around this transition point.

\begin{figure}
  \centering
  \includegraphics[scale=0.6]{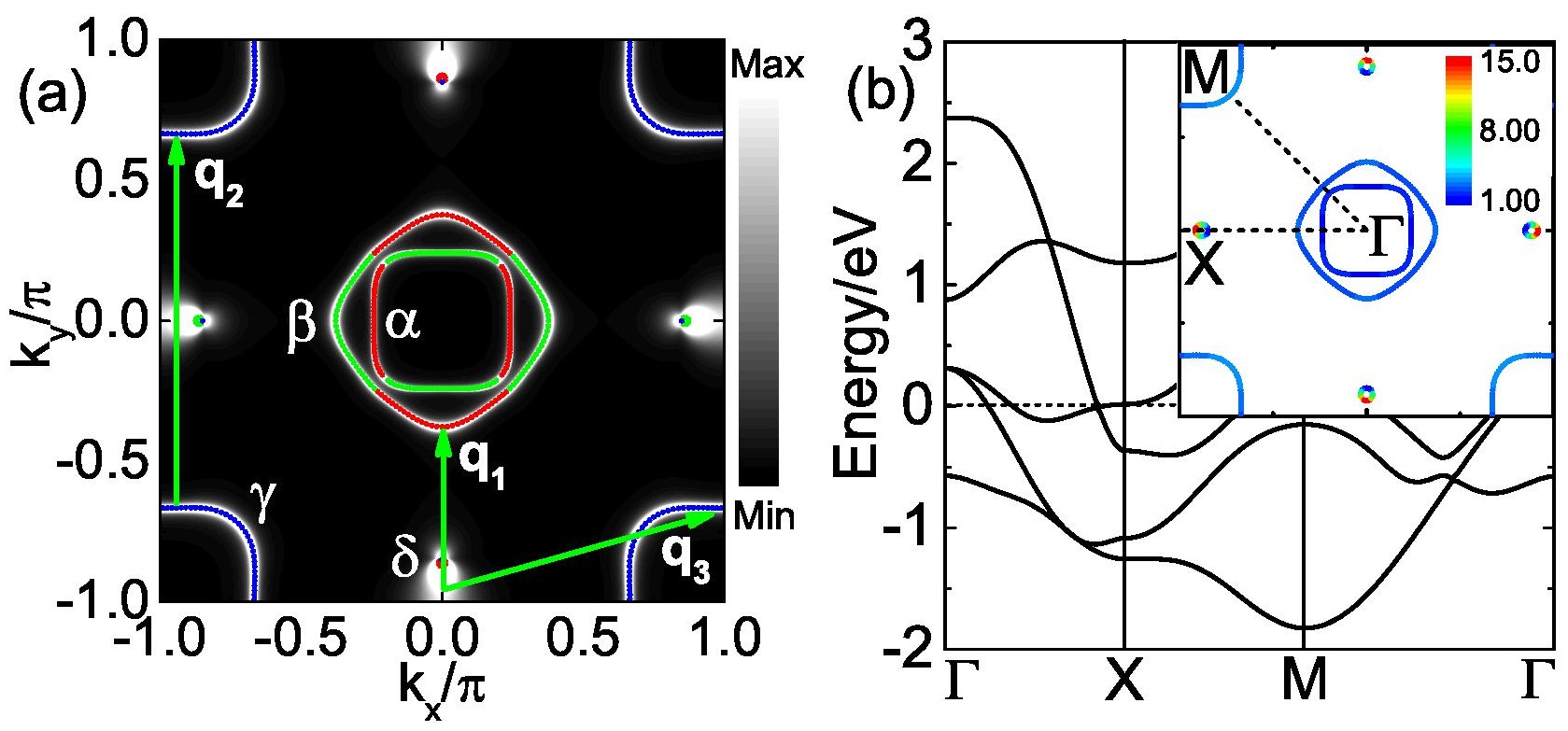}
  \caption{\label{fig1}(Color online) (a) FS in the unfolded BZ for KFe$_{2}$As$_{2}$ with $n=5.5$ at $k_{z}=0$.
  The dominant orbital weights along the FS are highlighted by colors(red: $d_{xz}$, green: $d_{yz}$, blue: $dxy$).
  The grayscale image in the background shows the intensity of the spectral function $A(\bm{k})$. The lines with an arrow indicate
  the transferred wave vectors for the dominant particle-hole scatterings. The four sets of FSs are labeled
  by $\alpha$, $\beta$, $\gamma$ and $\delta$, respectively. Here, the $\gamma$ FS corresponds to the outer FS in
  the folded BZ observed  by ARPES experiment according to its orbital characteristic. (b) Energy band structure. The dashed line
  indicates the Fermi level. The inset shows the density of state along the FS for $n=5.55$.}
\end{figure}

The model Hamiltonian consists of two parts:
\begin{equation}
H=H_{0}+H_{int}.
\end{equation}
$H_{0}$ is the
 effective five-orbital tight-binding model in the
unfolded Brillouin zone (BZ) developed by Graser \textit{et}
\textit{al}.\cite{SGraser2} to describe the energy band structure
of the BaK122. It reads
\begin{equation}
H_{0}=\sum_{\bm{k},\sigma}\sum^{5}_{\alpha,\beta=1}c^{\dag}_{\bm{k}\alpha\sigma}H_{\alpha\beta}(\bm{k})
c_{\bm{k}\beta\sigma},
\label{tight-binding}
\end{equation}
where $c^{\dag}_{\bm{k}\alpha\sigma}$ ($c_{\bm{k}\alpha\sigma}$)
creates (annihilates) an electron with spin $\sigma$ and momentum
$\bm{k}$ in the orbital $\alpha$. The details of
$H_{\alpha\beta}(\bm{k})$ can be found in
Ref. \onlinecite{SGraser2}. Fig. \ref{fig1}(a) shows the FS for the
electron concentration $n=5.5$ per Fe atom at $k_{z}=0$. The
relation between $n$ and $x$ is $n=6-0.5x$, so $n=5.5$ corresponds
to KFe$_{2}$As$_{2}$. Also shown in Fig. \ref{fig1}(a) is the
intensity map of the bare single-particle spectral function
$A(\bm{k})=-\mathrm{Im}[\mathrm{Tr}\hat{G}(\bm{k},\omega=0)]/\pi$
with $\hat{G}(\bm{k},\omega)$ the Green's function. We find that
the orbital characters of the FS are consistent with the
experimental results \cite{KOkazaki} and only a point-like FS
appears around the $X$ point. A characteristic of the electronic
structure is that there is a flat band (saddle point) close to the
Fermi level around $X$ [Fig. \ref{fig1}(b)], which results in four
Fermi patches (a region in the $k$ space with a spectral intensity
comparable to that on the FS) at $(0,\pm\pi)$ and $(\pm\pi,0)$ as shown in
Fig. \ref{fig1}(a). We note that recent ARPES experiment on heavily
over-hole-doped BK122 do observe similar large spectral intensity around $X$ \cite{NXu}.
The interactions between electrons
in $H_{int}$ are the standard multi-orbital on-site interactions
(see Appendix A), which include the intra-orbital (inter-orbital)
Coulomb interaction $U$ ($U^{\prime}$), the Hund's coupling $J$
and the pairing hopping $J^{\prime}$. In this paper, we choose
$U=0.6\mathrm{eV}$, $U^{\prime}=0.3\mathrm{eV}$ \cite{SGraser2},
and use the relations $J=J^{\prime}$ and $U=U^{\prime}+2J$.

Based on the scenario that the pairing interaction in the FeSCs arises from the exchange of spin and charge fluctuations, we can calculate the effective
electron-electron interaction using the RPA, which has been described in detail in the Appendix A. The singlet pairing interaction is given by
\begin{equation}
\hat{V}(q)=\frac{3}{2}\hat{U}^{s}\hat{\chi}^{s}(q)\hat{U}^{s}-\frac{1}{2}\hat{U}^{c}\hat{\chi}^{c}(q)\hat{U}^{c}
+\frac{1}{2}(\hat{U}^{s}+\hat{U}^{c}),
\label{eq-interaction}
\end{equation}
where $\hat{\chi}^{s}$ ($\hat{\chi}^{c}$) is the spin (charge) susceptibility and $\hat{U}^{s}$ ($\hat{U}^{c}$) is the interaction matrix for the spin (charge) fluctuation.

We confine our considerations to the dominant scattering occurring in the vicinity of the FS. The scattering amplitude of a Cooper pair from the state $(\bm{k},-\bm{k})$ on the FS $i$ to the state $(\bm{k}^{\prime},-\bm{k}^{\prime})$ on the FS $j$ is calculated from the projected interaction
\begin{align}
\Gamma_{ij}(\bm{k},\bm{k}^{\prime})\!=\!\!\sum_{\mu\nu\eta\varphi}&b^{\mu\ast}_{i}(-\bm{k})b^{\nu\ast}_{i}(\bm{k})\mathrm{Re}[V_{\varphi\nu,\mu\eta}(\bm{k}-\bm{k}^{\prime},\omega\!=\!0)] \nonumber \\
&\times b^{\eta}_{j}(\bm{k}^{\prime})b^{\varphi}_{j}(-\bm{k}^{\prime}),
\end{align}
where $b^{\mu}_{i}(\bm{k})=\langle\mu,\bm{k}|i,\bm{k}\rangle$ projects the band basis $|i,\bm{k}\rangle$ to the orbital basis $|\mu,\bm{k}\rangle$. Here, $i$ and $\mu$ are the band and orbital index respectively. We then solve the following eigenvalue problem:
\begin{equation}
-\sum_{j}\oint_{C_{j}}\frac{d\bm{k}^{\prime}_{\|}}{4\pi^{2}|\nabla E_{j}(\bm{k}^{\prime})|}\Gamma_{ij}(\bm{k},\bm{k}^{\prime})g_{j}(\bm{k}^{\prime})=\lambda g_{i}(\bm{k}),
\label{eigeq}
\end{equation}
where $E_{j}(\bm{k})$ is the energy of the tight-binding Hamiltonian (\ref{tight-binding}) for the band $j$ at the momentum $\bm{k}$ and $g_{i}(\bm{k})$ is the normalized gap function along the FS $i$. The integral in Eq. (\ref{eigeq}) is evaluated along the FSs. The most favorable SC pairing symmetry corresponds to the gap function with the largest eigenvalue $\lambda$. One merit of this method is that it can adequately include the effect of DOS on the FS, which is very important in BaK122.

To resolve the eigenequation (\ref{eigeq}), we use $256$ points along every hole-like FS and $16\sim128$ points along every electron-like FS depending on the size of the electron pocket. The temperature is set at $T=0.005\mathrm{eV}$,
and the calculation of the susceptibility is done with uniform $128\times128$ meshes.

\begin{figure}
  \centering
  \includegraphics[scale=0.58]{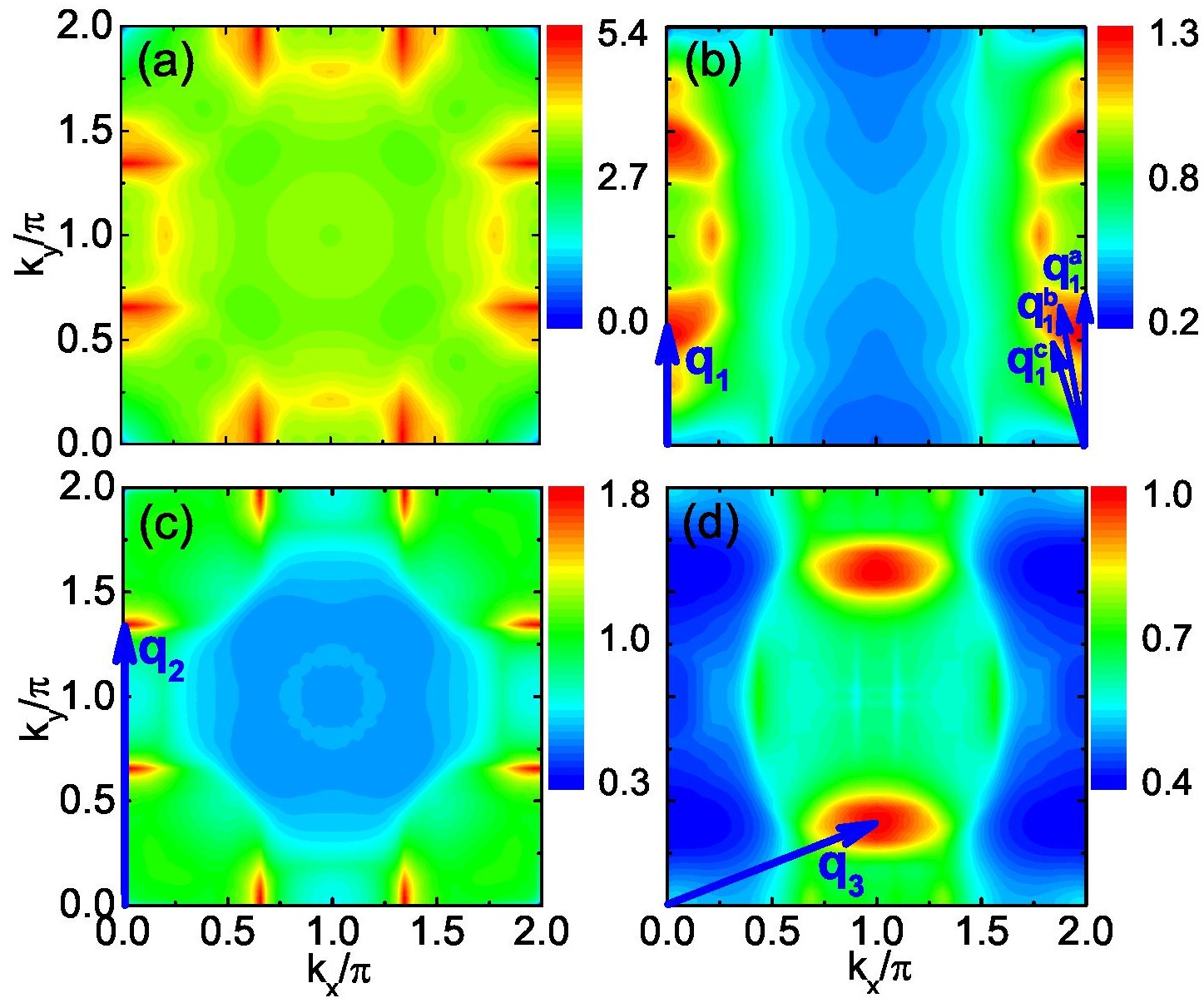}
  \caption{\label{fig2}(Color online) (a) Static physical spin susceptibility for $n=5.5$. (b), (c) and (d)
  are the components which contribute mainly to the static spin susceptibility: (b) $\chi_{\mu\mu,\mu\mu}$ with
  $\mu=d_{xz}$; (c) $\chi_{\mu\mu,\mu\mu}$ with $\mu=d_{xy}$; (d) $\chi_{\mu\nu,\mu\nu}$ with $\mu=d_{xz}$ and
  $\nu=d_{xy}$. The vectors indicate the positions of the peaks in spin susceptibilities. $q_{1}^{a}$,$q_{1}^{b}$
  and $q_{1}^{c}$ play basically the same role as $q_{1}$ due to
  the broad peak in $\chi$.
  }
\end{figure}
We will only discuss the spin fluctuations in the following because of the negligible role of charge fluctuation. In Fig. \ref{fig2}, we present the spin susceptibility
for $n=5.5$. Fig. \ref{fig2}(a) is the static physical spin
susceptibility
$\tilde{\chi}^{s}(\bm{q})=\sum_{\mu\nu}\chi^{s}_{\mu\mu,\nu\nu}(\bm{q},\omega=0)$,
which corresponds to that measured by the neutron scattering
experiments. It shows eight peaks located at the incommensurate
wave vectors $(\pi\pm0.34\pi,0)$ and their symmetric points, which
is consistent with the INS experiments on KFe$_{2}$As$_{2}$
\cite{CHLee,JPCastellan} and the previous theoretical result
\cite{KSuzuki}. We find that $\tilde{\chi}^{s}$ is governed primarily
by the intra-orbital spin fluctuations, and those within the
$d_{xz}$, $d_{yz}$ and $d_{xy}$ orbitals contribute mainly to it,
since the electronic states near the FS basically come from these
three orbitals [Fig. \ref{fig1}(a)]. In Figs. \ref{fig2}(b) and (c),
the intra-orbital spin fluctuations in the $d_{xz}$ (that in the
$d_{yz}$ orbital is rotated by $90$ degrees) and $d_{xy}$ orbitals
are presented. Usually, the spin fluctuations in the weak-coupling
scheme is resulted from the particle-hole scatterings of electrons
between the nesting FS. From the FS shown in Fig. \ref{fig1}(a),
one can find that there is no nesting condition for the
intra-$d_{xz}$-orbital spin fluctuation with the wavevector
$\bm{q}_{1}$ [Fig. \ref{fig2}(b)]. Instead, we find that it is the
scatterings of electrons between the $\beta$ FS and the Fermi
patches at $(0,\pm\pi)$[Fig. \ref{fig1}(b)] contribute essentially
to $\chi_{xz xz,xz xz}$. Whereas, the intra-$d_{xy}$-orbital spin
fluctuation with $\bm{q}_{2}$ [Fig. \ref{fig2}(c)] is resulted from
the nesting of the $\gamma$ FS as shown in Fig. \ref{fig1}(a). The
consequence of the Fermi-patch mechanism is that the peaks of
$\chi$ in the $d_{xz}$ orbital is much broader than those in the
$d_{xy}$ orbital. Another character is that in this doping level
$\bm{q}_{2}\approx2\pi-\bm{q}_{1}$, so the peaks of the
intra-orbital spin fluctuations in the $d_{xz}$ and $d_{xy}$
orbitals appear basically at the same wave vectors. Besides, we
find that the inter-orbital spin fluctuation
$\chi_{\mu\nu,\mu\nu}$ where $\mu=d_{xz}$ (or $d_{yz}$) and
$\nu=d_{xy}$ with the wavevector $\bm{q}_{3}$ also has a
comparable magnitude with the intra-orbital spin fluctuations as
shown in Fig. \ref{fig2}(d).
It mainly comes from the scatterings
of electrons between the $\gamma$ FS and the Fermi patches near
$(0,\pm\pi)$ as indicated by $\bm{q}_{3}$ in Fig. \ref{fig1}(a).
Though not showing up in the physical spin susceptibility,
this inter-orbital spin fluctuation is an important ingredient
in determining the pairing symmetry.

\begin{figure}
  \centering
  \includegraphics[scale=0.72]{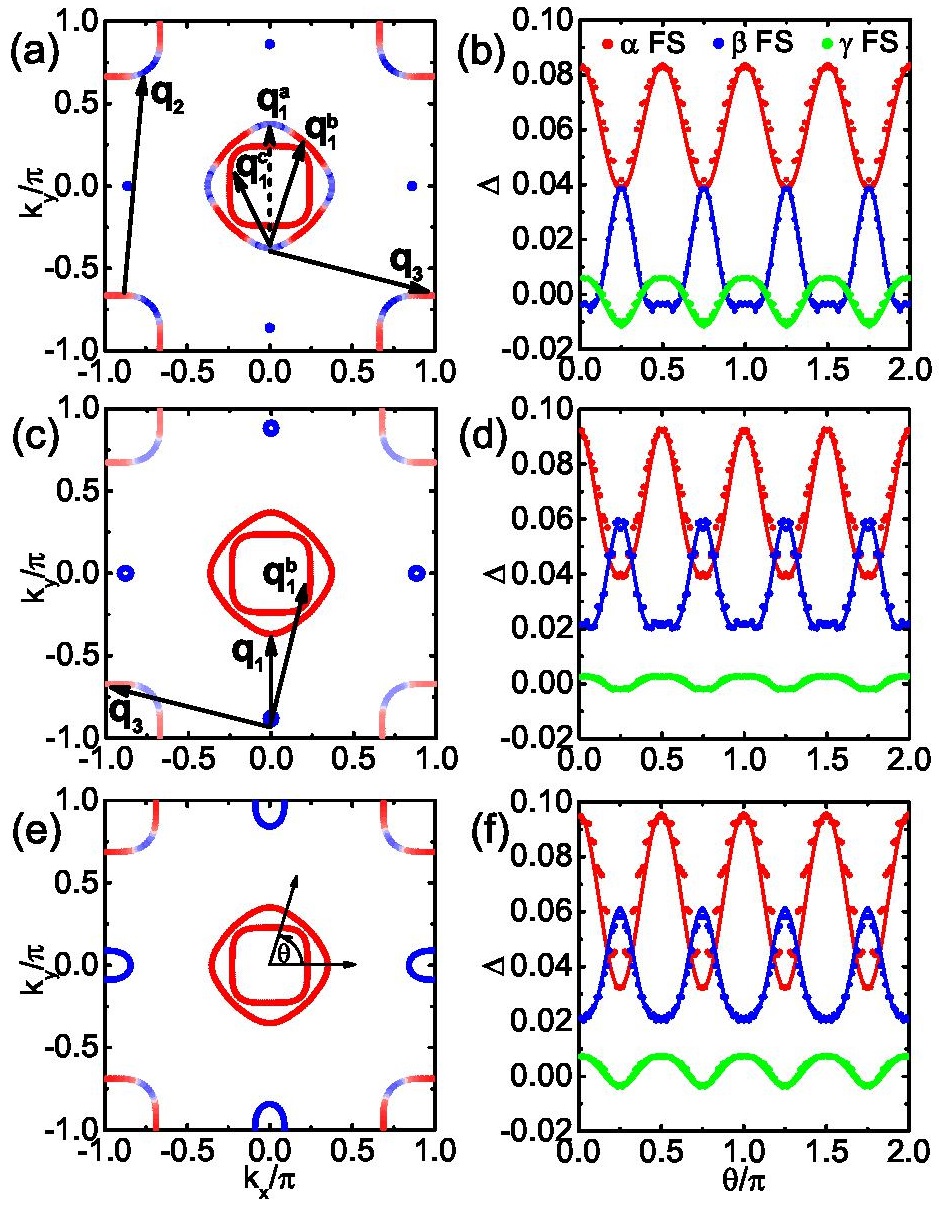}
  \caption{\label{fig3}(Color online) Dominant gap functions $g(\bm{k})$ and their sign structures along the FSs for different electron concentrations: (a) and (b) $n=5.5$, (c) and (d) $n=5.55$, (e) and (f) $n=5.63$. In (a), (c) and (e), the signs of $g(\bm{k})$ are shown by the following colors: red (positive) and blue (negative). The arrows in (a) and (c) indicate the typical vectors related to the peaks of the spin fluctuations. (b), (d) and (f) show $g(\bm{k})$ as a function of the angular $\theta$ indicated in (e). The fittings of $g(\bm{k})$ are shown as the solid lines[see text].}
\end{figure}
The dominant pairing functions $g(\bm{k})$ and their sign structures obtained from Eq. (\ref{eigeq}) for three different electron concentrations are shown in Fig. \ref{fig3}.
Figs. \ref{fig3}(a) and (b) show the results at $n=5.5$
corresponding to the case of KFe$_{2}$As$_{2}$. An
obvious feature is that the gap function exhibits an unusual
FS-dependent multi-gap structure. The gaps on the $\beta$ and
$\gamma$ FSs reveal an eightfold sign reversal
[Fig. \ref{fig3}(a)], whereas that on the $\alpha$ FS (the inner
pocket) is nodeless. In addition, the magnitude of the gap on the
$\gamma$ FS is much smaller than those on the $\alpha$ and $\beta$
FSs [Fig. \ref{fig3}(b)]. With a slight increase of the electron
concentration, we find that the gap anisotropy on these FS
drastically changes, as shown in Figs. \ref{fig3}(c-f) for $n=5.55$
($x=0.9$) and $n=5.63$ ($x=0.74$). The octet node structure on the
$\beta$ FS disappears completely, and the gaps on both the $\beta$
and $\alpha$ FSs are nodeless. While the gap on the $\gamma$ FS
still has the octet nodal structure. The obtained gap structure
for KFe$_{2}$As$_{2}$ and its doping dependence are consistent
with the recent laser ARPES results\cite{KOkazaki,YOta}.
Furthermore, the pairing functions on all three FSs
can be well fitted in an unified manner with the function
$g(k)=\Delta(a_{0}+a_{1}\cos4\theta+a_{2}\cos8\theta)$\cite{KOkazaki}, as shown
by the solid lines in Fig. \ref{fig3}(b), (d) and (f) with the
fitting parameters given in the Appendix B.

Whitin the spin-fluctuation mechanism, the pairing interaction in the
spin-singlet channel is positive (repulsive)[see
Eq. (\ref{eq-interaction})]. Thus, the most favorable SC gap should
satisfy the condition $g(\bm{k})g(\bm{k}+\bm{Q})<0$, where
$\bm{Q}$ is the typical wavevector at which the spin fluctuation
has a peak. As the $\delta$ FS is very tiny for $n=5.5$, it
doesn't play a role in determining the gap signs. According to the
general gap equation shown in the Appendix A, the
scattering of a Cooper pair mediated by the intra-orbital spin
fluctuation will happen between the FSs with the same orbital
character. Due to the orbital weights of the $\alpha$ and $\beta$
FSs shown in Fig. \ref{fig1}(a), three typical wavevectors
$\bm{q}_{1}^{a}$, $\bm{q}_{1}^{b}$ and $\bm{q}_{1}^{c}$, which
correspond to the broad peak of the intra-$d_{xz}$-orbital spin
fluctuation resulting from the scatterings of electrons in the
Fermi patches, connect the FS pieces with the $d_{xz}$ orbital
character. Due to the symmetry constraint for the spin singlet
pairing, the gap on the FS pieces connected by $\bm{q}_{1}^{a}$
can not change sign. While, those on the FS pieces connected by
$\bm{q}_{1}^{b}$ and $\bm{q}_{1}^{c}$ should change sign. Thus, it
leads to the anomalous gap structures on the $\alpha$ and $\beta$
FSs. This analysis is also applied to the sign change of
$g(\bm{k})$ within the $\beta$ FS connected by $\bm{q}_{2}$ which
is the characteristic wavevector of the intra-$d_{xy}$-orbital
spin fluctuation. While, the sign change between the $\beta$ and
$\gamma$ FSs connected by $\bm{q}_{3}$ is required by the
inter-orbital spin fluctuation as shown in Fig. \ref{fig2}(d).

With the increase of electron density, such as for $n=5.55$ and
$n=5.63$, the Fermi level will be pushed towards the flat band.
Consequently, the $\delta$ FS as well as the DOS on it will be
enlarged. Though we do not
find noticeable change in spin fluctuations correspondingly, the
essential change is that now the $\delta$ FS plays an important
role in determining the sign structure of $g(\bm{k})$. In
Fig. \ref{fig3}(c), we plot the typical vectors by which the gaps
on the connected FS change signs. The same situation happens for
$n=5.63$. In these cases, the gap
function on the $\beta$ FS is nodeless, while the sign structure
of $g(\bm{k})$ on the $\gamma$ FS remains the same as that for
$n=5.5$.

\begin{figure}
  \centering
  \includegraphics[scale=0.72]{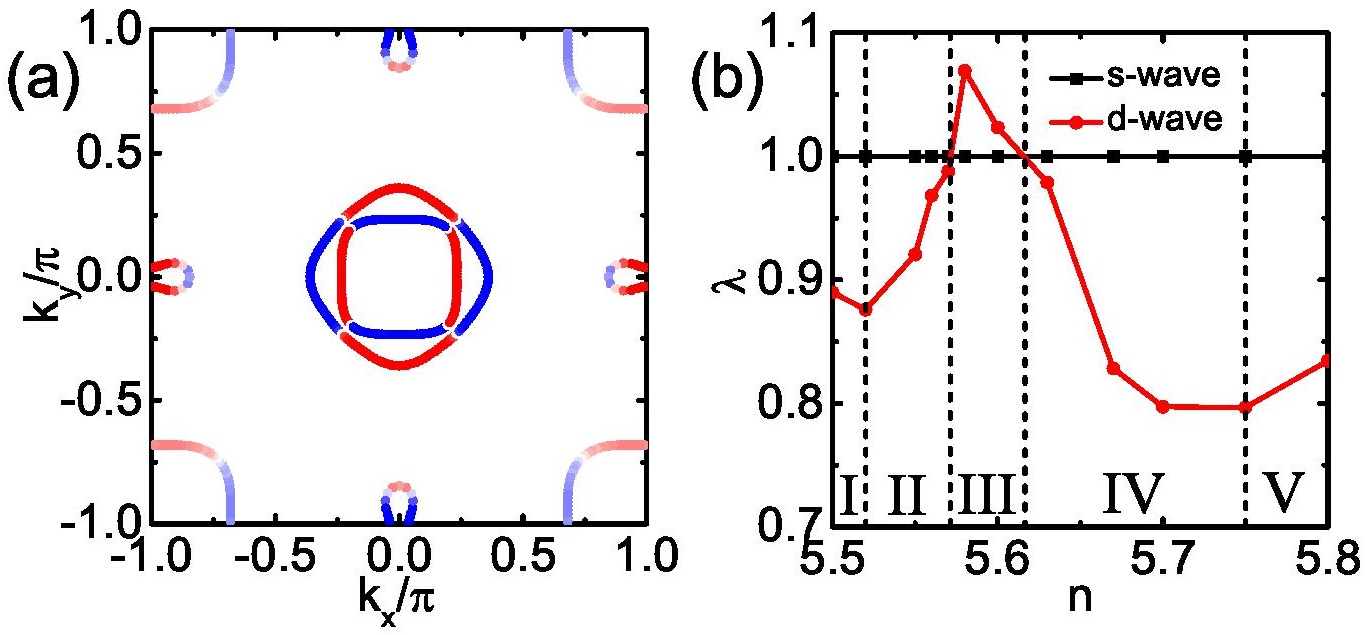}
  \caption{\label{fig4}(Color online) (a) The dominant gap function for $n=5.6$. (b) The largest eigenvalues $\lambda$ in the $s$-wave (normalized to 1) and $d$-wave channels as a function of the electron concentration $n$. The Roman numbers indicate the regions with different sign structures of the dominant gap function.}
\end{figure}
In fact, between $n=5.55$ and $n=5.63$, there is a Lifshitz
transition from the small off-$X$-centered hole FS pocket lobes to
that centered around the $X$ $(0,\pi)$ point, as can been seen
from a comparison between Fig. \ref{fig3}(c) and Fig. \ref{fig3}(e).
This Lifshitz transition has also been confirmed by the ARPES
experiment \cite{NXu}. Interestingly, we find that a $d$-wave
pairing state will prevail over the $s$-wave state around the
transition point, as seen from Fig. \ref{fig4}(a) where the most
favorable gap structure for $n=5.6$ is shown. The reason is that
the DOS on the $\delta$ FS near $X$ point is divergent at the
transition point, which makes the gap function of the $\delta$ FS changes its signs
between $(0,\pi)$ and $(\pi,0)$ points
according to Eq. (\ref{eigeq}). Besides this requirement, in fact,
the typical vectors by which the gaps on the connected FS change
signs are the same as those at $n=5.5$ shown in Fig. \ref{fig3}(c).
The only difference is that the sign structure is antisymmetric along the $\bm{k}_x$ and $\bm{k}_y$ directions in
this case, while it is symmetric at $n=5.5$ and $n=5.63$. Without
the addition requirement, in the latter case the nodeless gap on
both the $\alpha$ and $\beta$ FSs is favored energetically.

A detailed evolution of the gap symmetry with doping is presented in
Fig. \ref{fig4}(b), where the two leading eigenvalues of the gap
function Eq. (\ref{eigeq}) is shown (the optimal doping is
$n=5.8$). We can identify five regions according to the symmetry
and sign structure of the gap. In region (I), the gap is $s$-wave
with octet nodes on both $\beta$ and $\gamma$ FSs
[Fig. \ref{fig3}(a)]. In regions (II) and (IV), the gap is $s$-wave
with octet nodes only on the $\gamma$ FS [Fig. \ref{fig3}(c) and
(e)]. In region (III), the gap is $d$-wave [Fig. \ref{fig4}(a)]; In
region (V), the gap is nodeless $s_{\pm}$-wave and the peaks of
spin fluctuations are at $(0,\pm\pi)$ and $(\pm\pi,0)$
\cite{SGraser2}.
Considering the near degeneracy of the $s$ and $d$ wave, we
propose that the long sought $s+id$ pairing state in FeSC
\cite{WCLee,VStanev,CPlatt,MKhodas,SLYu,RMFernandes,FFTafti} would
be probably realized in the heavily overdoped BaK122 around the
Lifshitz transition point.

In summary, the pairing symmetry in the heavily overdoped Ba$_{1-x}$K$_{x}$Fe$_{2}$As$_{2}$ is studied based on the spin-fluctuation mechanism.
The exotic octet nodes of the superconducting gap and the unusual evolution of the gap with doping observed by the
recent experiments are explained in a unified manner.
The scattering of electrons related to the Fermi
patches is demonstrated to be mainly responsible for the incommensurate spin
fluctuations and consequently the gap structure. This Fermi-patch scenario provides a new viewpoint based on the large density of states resulting from the flat band, instead of the usual Fermi-surface-nesting picture where the Fermi surface topology is essential.

\begin{acknowledgments}
This work was supported by the National Natural Science Foundation of China (91021001, 11190023 and 11204125) and the Ministry of Science and Technology of China (973 Project Grants No.2011CB922101 and No. 2011CB605902).
\end{acknowledgments}

\section*{Appendix A: Random phase approximation (RPA) for multiorbital system}

The multiorbital Hubbard model we considered is given by
\begin{equation}
H=H_{0}+H_{int}.
\end{equation}
$H_{0}$ is the effective Hamiltonian without interactions, and $H_{int}$ can be written as
\begin{align}
H_{int}&=U\sum_{i\mu}n_{i\mu\uparrow}n_{i\mu\downarrow}+U^{\prime}\sum_{i,\mu<\nu}\sum_{\sigma\sigma^{\prime}}n_{i\mu\sigma}n_{i\nu\sigma^{\prime}} \nonumber \\
&+J\sum_{i,\mu<\nu}\sum_{\sigma\sigma^{\prime}}c^{\dag}_{i\mu\sigma}c^{\dag}_{i\nu\sigma^{\prime}}c_{i\mu\sigma^{\prime}}c_{i\nu\sigma} \nonumber \\
&+J^{\prime}\sum_{i,\mu<\nu}c^{\dag}_{i\mu\uparrow}c^{\dag}_{i\mu\downarrow}c_{i\nu\downarrow}c_{i\nu\uparrow},
\end{align}
where $n_{i\mu\sigma}=c^{\dag}_{i\mu\sigma}c_{i\mu\sigma}$ is the density operator at site $i$ of spin $\sigma$ in orbital $\mu$. $U$ ($U^{\prime}$) is the intra-orbital (inter-orbital) Coulomb interaction, $J$ and $J^{\prime}$ are the Hund's coupling and the pairing hopping respectively.

The effective pairing interaction mediated by spin and change fluctuations in the spin-singlet pairing channel is given by
\begin{equation}
\hat{V}(q)=\frac{3}{2}\hat{U}^{s}\hat{\chi}^{s}(q)\hat{U}^{s}-\frac{1}{2}\hat{U}^{c}\hat{\chi}^{c}(q)\hat{U}^{c}
+\frac{1}{2}(\hat{U}^{s}+\hat{U}^{c}),
\label{s_eq_1}
\end{equation}
where $\hat{\chi}^{s}$ ($\hat{\chi}^{c}$) is the spin (charge) susceptibility and $\hat{U}^{s}$ ($\hat{U}^{c}$) is the interaction matrix for the spin (charge) fluctuation. In RPA, the spin susceptibility $\hat{\chi}^{s}$ and charge susceptibility $\hat{\chi}^{c}$ are expressed as
\begin{equation}
\hat{\chi}^{s}(q)=[1-\hat{\chi}^{0}(q)\hat{U}^{s}]^{-1}\hat{\chi}^{0}(q)
\end{equation}
and
\begin{equation}
\hat{\chi}^{c}(q)=[1-\hat{\chi}^{0}(q)\hat{U}^{c}]^{-1}\hat{\chi}^{0}(q)
\end{equation}
respectively. The non-interacting susceptibility $\hat{\chi}^{0}$ is given by
\begin{equation}
\hat{\chi}^{0}_{\mu\nu,\eta\varphi}(q)=-\frac{T}{N}\sum_{k}G_{\eta\mu}(k+q)G_{\nu\varphi}(k)
\end{equation}
with the number of lattice sites $N$ and temperature $T$.  The Green's function is given by
\begin{equation}
\hat{G}(k)=[\mathrm{i}\omega_{n}-\hat{H}_{0}(\bm{k})+\mu]^{-1}.
\end{equation}
For an $m$-orbital system, the Green's function $\hat{G}$ is a $m\times m$ matrix, while the susceptibilities $\hat{\chi}^{0}$, $\hat{\chi}^{s}$, $\hat{\chi}^{c}$ and the interactions $\hat{V}(q)$, $\hat{U}^{s}$, $\hat{U}^{c}$ are $m^{2}\times m^{2}$ matrices.
In the above, $k\equiv(\bm{k},i\omega_{n})$ with $\omega_{n}=\pi T(2n+1)$. The interaction matrices $\hat{U}^{s}$ and $\hat{U}^{c}$ are given by:
\begin{equation}
U^{s}_{\mu\nu,\eta\varphi}=\left\{
                             \begin{array}{ll}
                               U, & \hbox{$\mu=\nu=\eta=\varphi$,} \\
                               J, & \hbox{$\mu=\nu\neq\eta=\varphi$,} \\
                               U^{\prime}, & \hbox{$\mu=\eta\neq\nu=\varphi$,} \\
                               J^{\prime}, & \hbox{$\mu=\varphi\neq\nu=\eta$,} \\
                               0, & \hbox{otherwise,}
                             \end{array}
                           \right.
\end{equation}
\begin{equation}
U^{c}_{\mu\nu,\eta\varphi}=\left\{
                             \begin{array}{ll}
                               U, & \hbox{$\mu=\nu=\eta=\varphi$,} \\
                               2U^{\prime}-J, & \hbox{$\mu=\nu\neq\eta=\varphi$,} \\
                               -U^{\prime}+2J, & \hbox{$\mu=\eta\neq\nu=\varphi$,} \\
                               J^{\prime}, & \hbox{$\mu=\varphi\neq\nu=\eta$,} \\
                               0, & \hbox{otherwise.}
                             \end{array}
                           \right.
\end{equation}

In the orbital representation, the superconducting gap equation (the ``Eliashberg" equation) is given by
\begin{eqnarray}
\lambda\Delta_{mn}(k)&=&-\frac{T}{N}\sum_{q}\sum_{\alpha\beta}\sum_{\mu\nu}
V^{s,t}_{\alpha m,n\beta}(q)G_{\alpha\mu}(k-q) \nonumber\\
& &\times G_{\beta\nu}(q-k)\Delta_{\mu\nu}(k-q).
\label{s_eq_2}
\end{eqnarray}
The most favorable superconducting pairing function is the eigenvector $\Delta_{mn}(k)$ with the largest eigenvalue $\lambda$.

Considering that the dominant scatterings of electrons occur in the vicinity of the Fermi surface (FS), we can reduce the effective interaction (\ref{s_eq_1}) and the ``Eliashberg" equation (\ref{s_eq_2}) to the FS. The scattering amplitude of a Cooper pair between two points at the FS [$(\bm{k},-\bm{k})\rightarrow(\bm{k}^{\prime},-\bm{k}^{\prime})$] is given by
\begin{align}
\Gamma_{ij}(\bm{k},\bm{k}^{\prime})\!=\!\!\sum_{\mu\nu\eta\varphi}&b^{\mu\ast}_{i}(-\bm{k})b^{\nu\ast}_{i}(\bm{k})\mathrm{Re}[V_{\varphi\nu,\mu\eta}(\bm{k}-\bm{k}^{\prime},\omega\!=\!0)] \nonumber \\
&\times b^{\eta}_{j}(\bm{k}^{\prime})b^{\varphi}_{j}(-\bm{k}^{\prime}),
\end{align}
where $b^{\mu}_{i}(\bm{k})=\langle\mu,\bm{k}|i,\bm{k}\rangle$ projects the band basis $|i,\bm{k}\rangle$ to the orbital basis $|\mu,\bm{k}\rangle$. Here, $i$ and $\mu$ are the band and orbital index respectively. In the calculations, we use the retarded Green's function and susceptibility by performing a Wick rotation $i\omega_{n}\rightarrow\omega+i\eta$. Then, the ``Eliashberg" equation (\ref{s_eq_2}) is reduced to
\begin{equation}
-\sum_{j}\oint_{C_{j}}\frac{d\bm{k}^{\prime}_{\|}}{4\pi^{2}|\nabla E_{j}(\bm{k}^{\prime})|}\Gamma_{ij}(\bm{k},\bm{k}^{\prime})g_{j}(\bm{k}^{\prime})=\lambda g_{i}(\bm{k}),
\label{s_eq_3}
\end{equation}
where $E_{j}(\bm{k})$ is the energy of the band $j$ at the momentum $\bm{k}$ and $g_{i}(\bm{k})$ is the normalized gap function along the FS $i$. The integral in Eq. (\ref{s_eq_3}) is evaluated along the FSs.

\section*{Appendix B: Fitting parameters of pairing functions}

The pairing functions on different FS from our calculations can be fitted by a unified function: $g_{i}(k)=\Delta_{i}(a_{i0}+a_{i1}\cos4\theta+a_{i2}\cos8\theta)$, where $i$ ($=\alpha$, $\beta$ or $\gamma$) represents one of the hole FSs. The fitting parameters for three typical electron concentrations $n=5.5$, $5.55$ and $5.63$ in Fig. 3 of the main text are list in Table \ref{table1}. Though the $\cos4\theta$ term dominates the octet-node structure, the $\cos8\theta$ term also plays an important role, especially for the $\beta$ Fermi surface at $n=5.5$.
\begin{table}[h]
\caption{\label{table1}
Parameters in fitting the pairing functions.
}
\begin{ruledtabular}
\begin{tabular}{ccccc}
\textrm{n}&
\textrm{FS}&
\textrm{$a_{0}$}&
\textrm{$a_{1}$}&
\textrm{$a_{2}$}\\
\colrule
5.5 & $\alpha$ & 0.061 & 0.023 & 0\\
5.5 & $\beta$ & 0.01 & -0.021 & 0.008\\
5.5 & $\gamma$ & 0.0015 & 0.009 & -0.001\\
\colrule
5.55 & $\alpha$ & 0.063 & 0.027 & 0.003\\
5.55 & $\beta$ & 0.034 & -0.0185 & 0.007\\
5.55 & $\gamma$ & 0.0005 & 0.0025 & 0\\
\colrule
5.63 & $\alpha$ & 0.064 & 0.032 & 0\\
5.63 & $\beta$ & 0.036 & -0.02 & 0.005\\
5.63 & $\gamma$ & 0.0022 & 0.0057 & -0.0005\\
\end{tabular}
\end{ruledtabular}
\end{table}

\end{document}